\documentclass[conference,letterpaper]{IEEEtran}

\usepackage{bsymb}
\usepackage{amssymb}
\usepackage{amsmath}
\usepackage{hyperref}
\usepackage{varioref}
\usepackage{xspace}
\usepackage{graphicx}
\usepackage{multirow}
\usepackage{paralist}
\usepackage[pdftex]{color}

\renewcommand{\Bool}{\mathbb{B}}
\renewcommand{\True}{\mathbf{T}}
\renewcommand{\False}{\mathbf{F}}

\newcommand{\assign}[1]{{\textbf{\textit{Assignment: }}}\textbf{{#1}}\\}
\newcommand{\condition}[1]{\\{\textbf{\textit{Condition: }}}\textbf{{#1}}\\\\}
\newcommand{\conditionNS}[1]{\\{\textbf{\textit{Condition: }}}\textbf{{#1}}\\}
\newcommand{\actionM}[2]{\mbox{}\\{\textbf{\textit{Action(s): }}}{#1} $:=$ {#2}\\}
\newcommand{\actionMNS}[2]{\mbox{}\\{\textbf{\textit{Action(s): }}}{#1} $:=$ {#2}}

\newcommand{\kG}[1]					{\textbf{\textsf{#1}}}
\newcommand{\kB}[1]					{\mathsf{#1}}

\newcommand{\kwSYSTEM}			{MACHINE}

\newcommand{\kwVARIABLES}		{VARIABLES}
\newcommand{\kwINVARIANT}		{INVARIANT}
\newcommand{\kwINITIALISATION}	{INITIALISATION}
\newcommand{\kwEVENTS}			{EVENTS}

\newcommand{\kwMEND}				{END}

\newcommand{\kwANY}				{any}
\newcommand{\kwWHERE}			{where}
\newcommand{\kwTHEN}			{then}

\newcommand{\kwSEES}			{SEES}

\newcommand{\kwEND}				{end}

\newcommand{\EventBSystem}[2] {
	{\footnotesize{
	$
	\begin{array}{@{}l}
		\kG{\kwSYSTEM} ~ \kB{#1} \\
		\quad \begin{array}{@{}l} #2 \end{array} \\
		\kG{\kwMEND}
	\end{array}
	$
	}}
}

\newcommand{\EventBSees}[1] {
	\kG{\kwSEES} ~ \kB{#1} \\
}

\newcommand{\EventBVarsC}[1] {
	\kG{\kwVARIABLES} ~ #1 \\
}

\newcommand{\EventBInvC}[1] {
	\kG{\kwINVARIANT} ~ #1 \\
}

\newcommand{\EventBInitC}[1] {
	\kG{\kwINITIALISATION} ~ #1 \\
}

\newcommand{\EventBEvents}[1] {
	\kG{\kwEVENTS} \\
	\quad \begin{array}{lll} #1 \end{array} \\
}

\newcommand{\EventBAny}[4] {
  \kB{#1} & = 	& 	\kG{\kwANY} ~ #2 ~ \kG{\kwWHERE} \\
			  && 	\quad \begin{array}{l} #3 \end{array} \\
			  && 	\kG{\kwTHEN} \\
			  && 	\quad \begin{array}{l} #4 \end{array} \\
			  && 	\kG{\kwEND} \\
}

\newenvironment{spec}
{
\begin{tabular*}{.45\textwidth}{c}
\hline
\end{tabular*}
}
{
\\
\begin{tabular*}{.45\textwidth}{c}
\hline\\
\end{tabular*}
}

\begin{document}

\title{Towards a Formalism-Based \\ Toolkit for Automotive Applications}





\author{
	Rainer Gmehlich, Katrin Grau, Felix Loesch\\
	Robert Bosch GmbH, Stuttgart, Germany
\and
	Alexei Iliasov, Michael Jackson, Manuel Mazzara\\
	School of Computing Science, Newcastle University, UK
}

%

\IEEEcompsoctitleabstractindextext{%

}

\maketitle

\begin{abstract}
The success of a number of projects has been shown to be significantly improved by the use of a formalism . However, there remains an open issue: to what extent can a development process based on a singular formal notation and method succeed. The majority of approaches demonstrate a low level of flexibility by attempting to use a single notation to express all of the different aspects encountered in software development. Often, these approaches leave a number of scalability issues open. We prefer a more eclectic approach. In our experience, the use of a formalism-based toolkit with adequate notations 
for each development phase is a viable solution. Following this principle, any specific notation is used only where and when it is really suitable and not necessarily over the entire software lifecycle. The approach explored in this article is perhaps slowly emerging in practice ---we hope to accelerate its adoption. However, the major challenge is still finding the best way to instantiate it for each specific application scenario. In this work, we describe a development process and method for automotive applications which consists of five phases. The process recognizes the need for having adequate (and tailored) notations (Problem Frames, Requirements State Machine Language,  and Event-B) for each development phase as well as direct traceability between the documents produced during each phase. This allows for a step-wise verification/validation of the system under development. The ideas for the formal development method have evolved over two significant case studies carried out in the DEPLOY project.

\end{abstract}

\section{Introduction}
\label{sec:introduction}

\IEEEPARstart{}One of the lessons of the DEPLOY project \cite{DEPLOY} is that the industrial application of formal modelling cannot fully succeed by employing just one notation, paradigm and methodology. The focus of academic research in DEPLOY was a method called Event-B\cite{EventBBook} - a general-purpose, event-based and refinement driven formal modelling method - and a toolkit supporting it - the Rodin Platform \cite{RodinPlatform}. A wide range of case studies and formal developments were conducted during the predecessor RODIN project \cite{RODIN} and this provided a reasonable expectation that the method and the Platform could succeed in large-scale industrial deployments. This paper describes the experience of deploying Event-B in an automotive sector. In particular, we tell the story of how a sole notation approach based on Event-B has gradually transformed itself into a rich assembly of diverse notations and techniques.

Formal/mathematical notations have existed for a long time and have been used
to specify and verify systems. Examples are process algebras 
(a short history by Jos Baeten in \cite{BaetenPA}), specification languages
like Z (early description in \cite{AbrialZ-book}), B \cite{AbrialB-book} and 
Event-B \cite{Abrial10}. The Vienna Development Method (VDM) is one of 
the earliest attempts to establish a formal method for the development of 
computer systems \cite{BjornerVDM78,Jones80a,Jones90a}. A survey of these (and others) 
formalisms can be found in \cite{Mazzara10} while a discussion on
the methodological issues of a number of formal methods is presented in
\cite{MazzaraDEPLOY09, Mazzara11, WoodcockEt09}.

All these approaches (and others described in the literature) still leave 
an open issue, i.e.,~they are built around strict formal notations 
which affect the development process from the very beginning.
These approaches demonstrate a low level of flexibility. It is indeed
not reasonable to expect that a single notation can express all 
the different aspects encountered during the software development cycle. 
Therefore, these methods seem to work only for small problems, leaving 
a number of scalability issues open.

In this paper, following the experience accumulated during the FP7 DEPLOY project 
\cite{DEPLOY}, we offer to consider a position where a development toolkit and its supporting method are based on a range of notations 
or formalisms that complement each other. Each such notation may target a specific stage of a development - early requirements elicitation, concrete requirements, abstract design, concrete design, first prototype, final product and so on - with some degree of overlap to give a degree of confidence when progressing through notations. 

One challenge is in making a seamless methodological connection between diverse notations and methods: how to ensure that results expressed in one notation are carried over to the next stage based on a differing notation without misinterpretation of specification statements and mis-attribution of validation results. 

Another challenge is in the provision of a modelling environment that can adopt all such notation and possibly integrate validation tools behind them. Finally, and perhaps most importantly, it is important to have some degree of notational flexibility - industrial software development is already based on long tool chains and the most likely path to succeed for a formal development method to succeed is to morph into such chain. That is, be prepared to work with new input and output notations.

\section{Background}
\label{sec:background}
This section gives a short introduction to the different methods on which the approach proposed is based. It is not intended to be exhaustive, but it provides the reader with relevant pointers for further investigation.


\subsection{Problem Frames}
\label{subsec:background_pfa}

PFA \cite{Jackson00} focuses on systems in which the computer interacts with the physical world to achieve a required behaviour there. Stakeholders in the system --users, sponsors, operators, regulators and others-- want this behaviour to satisfy certain properties. These desired properties may be expressed in various forms and with various degrees of exactness: for an industrial press a vital desire is the operator's safety; for an electronic purse system it is  conservation of money in any transaction between two purses even if the transaction fails or is aborted. The requirements engineer must understand these desires and design a feasible joint behaviour of the computer and the world that will satisfy them. 

In PFA this task is understood in terms of three principal parts. First, the machine: this is the computer executing the software that will eventually be developed. Second, the problem world, seen as an assemblage of distinct domains interacting with each other and with the computer. Third, the system requirement, initially seen as the set of desired properties of the system behaviour. The system is represented in a problem diagram. The diagram shows the computer, the problem domains, and the interfaces of shared phenomea at which they interact; the requirement is represented by a distinguished block linked to the problem domains to whose phenomena it refers. The requirements engineering task is to specify the given properties $W$ of the problem world domains, the behaviour $M$ of the computer, and the required joint behaviour $R$ resulting from their interactions. The entailment $M,W \vDash R$ must hold, and the behaviour $R$ must exhibit the properties desired by the stakeholders. 

For a realistic system $M, W, R$ and the desired properties will be complex. The problem is therefore decomposed into subproblems, each represented by a problem diagram. A subproblem is a closed independent projection of the original problem, ignoring all interactions with other subproblems. Recombination is deferred until each subproblem is well enough understood in isolation. A further task is then to design the temporal composition of the subproblem behaviours and to resolve any interference and conflict arising in their resulting interactions. 

This specification of system behaviour does not map directly either to an Event-B specification or to a software architecture: refactoring is a further step in the path to implementation. It is a fundamental claim of PFA that the cost of this refactoring is amply compensated by the clarity that can be achieved in the requirements engineering task itself and  the consequent improvement in system quality and dependability. 

\subsection{Requirements State Machine Language (RSML)}
\begin{table}[t]
\begin{center}
\caption{Example of an AND/OR table}
\label{tab:rsml_andortable}
\begin{tabular}{|l||l|l|l|} \hline
X $>$ Y & $\True$ & $\False$ & $\bullet$ \\ \hline
A $<$ B & $\True$ & $\False$ & $\bullet$ \\ \hline
S = PRESSED & $\bullet$ & $\True$ & $\True$ \\ \hline
Y = ON & $\bullet$ & $\bullet$ & $\True$ \\ \hline
\end{tabular}
\end{center}
\end{table}

\label{subsec:background_rsml}
The Requirements State Machine Language (RSML) \cite{Leveson:1994} is a formal black-box specification language invented by Nancy Leveson and has been widely applied in the avionic industry for the specification of complex state-based embedded systems like the transition collision avoidance system (TCAS II). RSML was developed in order to have precise description of the functional behaviour of state-based systems which is formal enough to reason about general aspects like completeness and consistency of state machines \cite{Heimdahl:1996} but still easy enough to be understandable by engineers. The language itself consists of concepts for structuring a large specification, i.e., the language supports modules with defined interfaces as well as formal concepts for describing state machines based on statecharts \cite{Harel:1987} extending state diagrams with state hierarchies and broadcast communications. 

An important concept introduced by RSML is the concept of AND/OR tables which are used to describe conditions for state transitions and conditions for the assignment of variables. Table \ref{tab:rsml_andortable} shows an example for an AND/OR table.  The far-left column of the AND/OR table lists the logical phrases. Each of the other columns is a conjunction (logical AND) of those phases and contains the logical values of the expressions. If one of the columns is true, then the table evaluates to true. A column evaluates to true if all of its elements match the truth values of the associated predicates. A dot denotes "don't care".

\subsection{Event-B}
\label{subsec:background_eventb}

The Event-B Modelling Language \cite{Abrial10} was developed 
by J.-R. Abrial and his team at ETHZ as a specialization of the 
B-Method \cite{AbrialB-book} and it is used to describe formally 
systems and reason mathematically about their properties.

An Event-B development starts with the creation of a relatively abstract specification. A cornerstone of the Event-B method is the stepwise development that facilitates a gradual design of a system implementation through a number of correctness-preserving \emph{refinement} steps. The general form of an Event-B model (or \emph{machine}) is shown in Figure \ref{EventB_structure}. Such a model encapsulates a local state (program variables) and provides operations on the state. The actions (called \emph{events}) are characterised by a list of local variables (parameters) $vl$, a state predicate $g$ called \emph{event guard}, and a next-state relation $S$ called \emph{substitution} or event \emph{action}.

\begin{figure}[t]
   \centering
\EventBSystem{M}{
	\EventBSees{Context}
	\EventBVarsC{v}
	\EventBInvC{I(c, s, v)}
	\EventBInitC{R(c, s, v')}
	\EventBEvents{
		\EventBAny{E_1}{vl}{g(c, s, vl, v)}{S(c, s, vl, v, v')}
		\dots \\
	}
}
   
  \caption{Event-B machine structure.}
  \label{EventB_structure}
\end{figure}


Event guard $g$ defines the condition when an event is \emph{enabled}. Relation $S$ is given as a generalised substitution statement \cite{BBook} and is either deterministic ($x \bcmeq 2$) or non-deterministic update of model variables. The latter kind comes in two notations: selection of a value from a set, written as $x \bcmin \{2, 3\}$; and a relational constraint on the next state $v'$, e.g., $x \bcmsuch x' \in \{2, 3\}$.

\par The {\footnotesize{\kG{\kwINVARIANT}}} clause contains the properties of the system, expressed as state predicates, that must be preserved during system execution. These define the \emph{safe states} of a system. In order for a model to be consistent, invariant preservation is formally demonstrated. Data types, constants and relevant axioms are defined in a separate component called \emph{context}.

Model correctness is demonstrated by generating and discharging \emph{proof obligations} - theorems in the first order logic. There are proof obligations for model consistency and for the refinement link - the forward simulation relation - between the pair of \emph{abstract} and \emph{concrete} models. One can say that an abstract model serves as a contract when realising the concrete model.

\section{Our Approach}
\label{sec:approach}


Our approach to formal development process for automotive applications \footnote{Automotive applications contain discrete and continuous parts (closed loop controllers). In the case studies we concentrated on the discrete part of the system. We decided not to model the continuous part and only used an abstract notion of time.} evolved after having experimented over two significant case studies in the DEPLOY project \cite{DEPLOY}. During these case studies we found that the semantic gap between informal descriptions (i.e., requirements in natural language and formal descriptions of the system in Event-B) is significant. Checking that the formal descriptions are consistent with the informal descriptions turned out to be a difficult task because of the inevitable vagueness of informal descriptions and missing traceability links between the informal and formal descriptions. 

In order to bridge this gap and to progress incrementally from an informal to a formal description of the system, our approach consists of five phases (requirements, specification, formal modelling, formal verification, and code generation) in each of which a carefully selected notation and methods are used. A specific formalism is used only where and when it is really suitable and not over the complete development cycle. 

The outcome of each phase during the development process is an adequate document which describes the results of each phase and which can be used to communicate with other stakeholders like managers, customers, and other developers during the development process. To ensure traceability between the documents produced in different phases we establishment and maintain links between requirements and modelling artefacts implementing them. Hence, for instance, an RSML and Event-B models may be related by comparing model elements related to same requirements. Since each notation is tailored to its phase it becomes much easier to do informal model validation - ensuring that the model adequately addresses informal requirements. This process is crucial at the stage when concrete requirements are elaborated. 

\begin{figure}[t]
\centering
\includegraphics[width=0.5\textwidth]{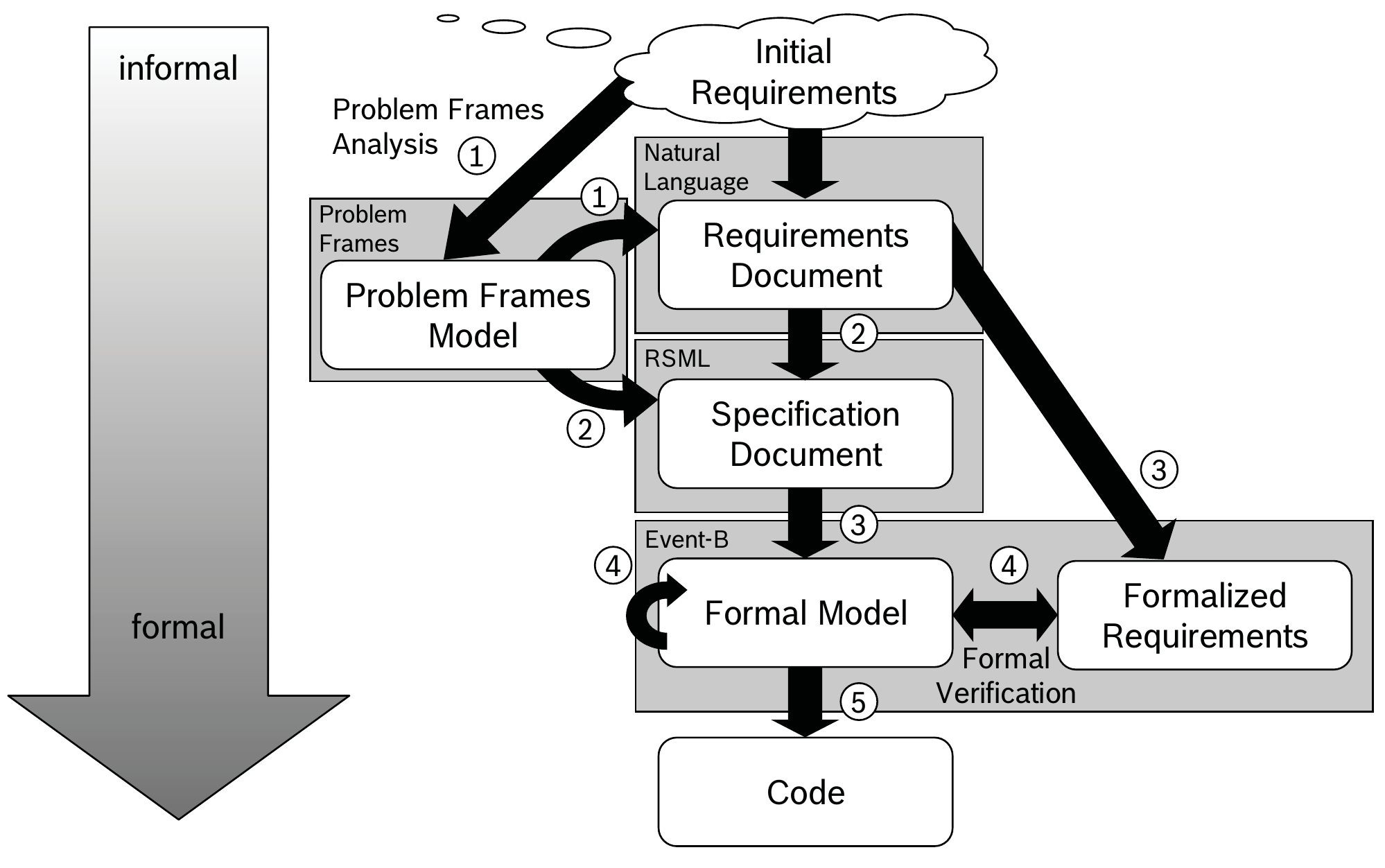}
\caption{Formal development process}
\label{fig:development_process}
\end{figure}

Figure \ref{fig:development_process} graphically depicts our development process. 

The starting point for our process is an abstract idea of the system and/or some vague initial requirements (see step 1 in Figure \ref{fig:development_process}). In order to produce a requirements document for further development these vague initial requirements have to be analyzed and concrete requirements have to be developed. For this analysis and development PFA \cite{Jackson00} is applied. The main outcome of this phase is a requirements document in natural language which also contains assumptions on the system to be developed. We deliberately chose to describe requirements in natural language in order to make it as easy as possible to discuss the requirements with different stakeholders in the development process (e.g., customers and managers). Section~\ref{subsec:method_requirements} describes this phase.

The next phase in the process is the specification phase in which the desired functional behaviour of the system and the architecture is described in a precise way using RSML \cite{Leveson:1994} (see step 2 in Figure \ref{fig:development_process}). Inputs to this phase are the requirements document as well as the Problem Frames model. The outcome of the specification phase is a specification document which contains a description of the architecture of the solution as well as a detailed description of the functional behaviour of each component. A detailed description of this phase is presented in Section \ref{subsec:method_specification}.

After the specification phase the formal modelling phase follows (see step 3 in Figure \ref{fig:development_process}). During this phase the specification is translated into a formal model written in a formal language (e.g., Event-B). The main activities in this phase are the formalization of the functional behaviour, i.e., how the system is achieving it, as well as the formalization of requirements, i.e., what the system should do. Section \ref{subsec:method_formal_modelling} contains a detailed description of the formal modelling phase. 

The next phase in our approach is the formal verification phase (see step 4 in Figure \ref{fig:development_process}). In this phase the refinements of the formal model as well as the formalized requirements are verified on the model using formal verification techniques such as theorem proving and model checking. The outcome of the verification phase is a verified formal model with regard to the formalized requirements.

\section{Formal Development Method}
\label{sec:method}
This section presents a detailed description of the five phases of our approach. For each phase of our approach the specific requirements and constraints
for choosing an adequate (formal) notation are discussed before we present the arguments for how the chosen notation fulfils the requirements of each phase. The description 
of each phase is illustrated using an example from our second case study in the DEPLOY project \cite{DEPLOY}.

The system we analyzed in our second case study was a Start/Stop System which automatically stops the engine, e.g., at traffic lights, to save fuel (see also~\cite{DEPLOYD38}). The engine will be automatically restarted when the driver wants to move the car again. The system is an embedded real time system. However, contrary to other software functions in the automotive domain, the Start/Stop System only consists of discrete functionality containing a complicated state machine for determining when to stop and when to start the engine.

%

\subsection{Requirements Development}
\label{subsec:method_requirements}


\textbf{Constraints}
The starting point for a new product or a new feature for a product is usually an abstract idea, some vague initial requirements. To produce a requirements document which can be used for further development these initial requirements have to be refined. We used Problem Frames \cite{Jackson00} for the problem analysis and the central idea of this first part of the development process is to concentrate on the problem that has to be solved, not on possible solutions. During this analysis, having some structure helps to find a systematic way to analyse the problem. On the other hand, a completely formal notation would restrict the freedom needed during this early phase.\\



\textbf{Description of method}
In PFA we start with an abstract diagram, an overview of the world of which the system to be built is a part. An abstract requirement describes the effect the system has on the world. Note that the requirement does not refer to the system itself (which would be a restriction of the solution). After this abstract examination more concrete sub-problems are considered. In these sub-problems one aspect of the overall problem is developed in detail with requirements that refer only to this specific aspect. The problem of how to recombine these different aspects is postponed and addressed after the development of all sub-problems. In every sub-problem there is at least one requirement. This requirement refers only to the sub-problem. After the development of every sub-problem in isolation the recombination must address the prioritization of the single sub-problems.

\begin{figure}[ht!]
\centering
\includegraphics[width=0.49\textwidth]{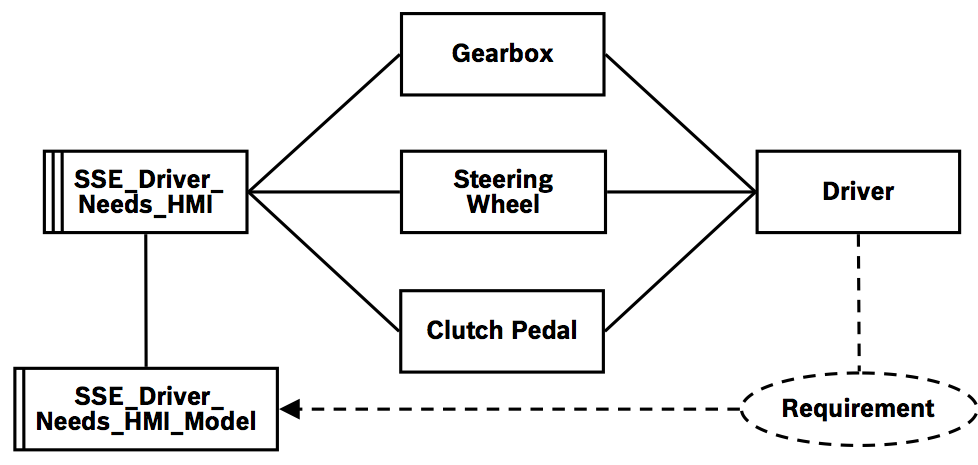}
\caption{Problem Frames sub-problem}
\label{fig:DriverNeedsHMI}
\end{figure}

\textbf{Example:}
The Start/Stop System is not allowed to prevent the driver from moving the car whenever he or she wishes to do so.
This aspect of the Start/Stop System is treated in the Problem Frames sub-problem shown in Figure~\ref{fig:DriverNeedsHMI}. The machine, i.e., the box with the double vertical stripe, is called $\mathrm{SSE\_Driver\_Needs\_HMI}$, referring to the fact that this subproblem concentrates of the needs of the driver, which are deduced by the HMI (Human-Machine Interface). To be able to solve the recombination problem the engine is not part of the subproblem. Instead a designed domain called $\mathrm{SSE\_Driver\_Needs\_HMI\_Model}$ is used and therefore the requirement does not refer to the engine (as in the requirements document) but to this designed domain, i.e., the box with the single vertical stripe. The designed domain has a phenomenon named $\mathrm{HMI\_Stop\_Ena}$ (there is another phenomenon called $\mathrm{HMI\_Strt\_Req}$, which is not relevant for this example but will be used in Example 4). The phenomenon stores the information of this subproblem related to the stopping of the engine, i.e., of whether this sub-problem enables the Start/Stop System to stop the car or not. For more details please see~\cite{DEPLOYD38}. In the domain $\mathrm{Driver}$ a model of the driver is defined, which states the connection mentioned in example 1 in 1.) and 2.) between the wishes of the driver and the steering wheel, the clutch and the gearbox. The steering wheel can be used or not used, the clutch pedal can be pressed or released, the gearbox can be in neutral or not in neutral.

\textbf{Summary}
The use of Problem Frames helps to concentrate on the problem to solve and develop a better understanding of how the system to build is supposed to affect the surrounding world. The additional requirements document in natural language is the basis for discussions with all stakeholders.


\subsection{Specification}
\label{subsec:method_specification}
\textbf{Constraints} 
After having produced a requirements document containing the requirements and assumptions on the system, the next step is to develop a detailed specification which should include a precise description of the functional behaviour of the system as well as a formalisation of the overall architecture. 


A specification method must be understandable by engineers who are not familiar with formal notations like Event-B. RSML \cite{Leveson:1994} is ideally suited for our task of specifying the functional behaviour of state-based automotive systems because it is easy enough to be understandable for engineers but still formal enough to reason about general aspects of state-based systems and fulfils the other constraints described above. The outcome of the specification phase is a specification document written in RSML which is then used as input for the formal modelling phase.

\textbf{Description of method}
For the specification of the system we start with the requirements document and the Problem Frames model produced during the requirements development phase. These documents contain requirements and assumptions about the system to be developed but do not contain a precise description of the desired functional behaviour of the system. Thus, the task for the specification phase is to specify the desired functional behaviour such that it fulfils the set of requirements described in the requirements document. In order to structure the solution, the first step during specification is to think about the general architecture of the system. 
As with the decomposition of the problem in the requirements development phase, the solution is decomposed into components that describe specific aspects. For each component its interface is precisely defined using typed input and output variables. Components communicate with other components via shared variables, e.g., the output variables of component $A$ serve as input variables to component $B$ and vice versa. If necessary, a component may also contain internal variables to store values derived from input variables. 

\begin{figure}[ht!]
\centering
\includegraphics[width=0.45\textwidth]{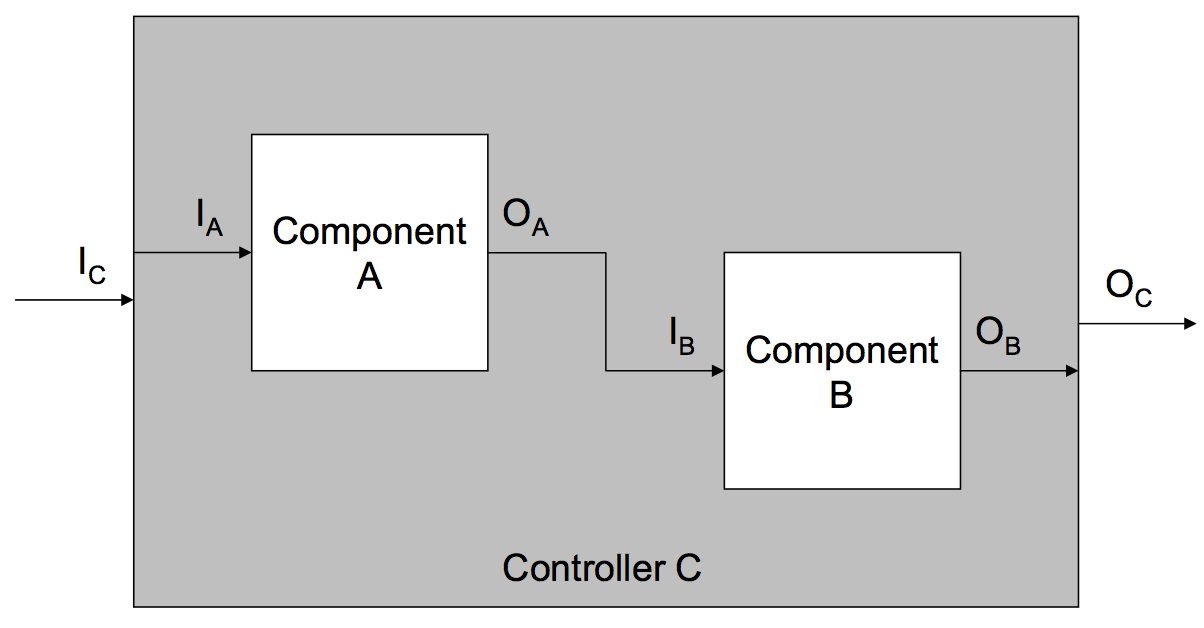}
\caption{RSML - Static structure with components and interfaces}
\label{fig:rsml_static_structure}
\end{figure}

Figure \ref{fig:rsml_static_structure} shows an exemplary static structure of an embedded controller
consisting of two components $A$ and $B$ and their interfaces.


The desired functional behaviour of a component is specified using two concepts. The first concept --called assignment specification-- is to relate output variables directly with conditions on input variables using AND/OR tables. The following example taken from the Start/Stop System case study illustrates this concept. 

\textbf{Example 3:} The value of the boolean output variable $\mathrm{HMI\_Stop\_Ena}$ is dependent on specific conditions on the input variables $\mathrm{Clutch\_Pedal, Steering\_Wheel}$ and $\mathrm{Gearbox}$. These conditions are specified by the assignment specification shown in Figure~\ref{fig:rsml_assignment_spec}.   
\begin{figure}[ht!]
\begin{spec}
\assign{HMI\_Stop\_Ena}
\condition{$d$}
  \begin{tabular*}{.45\textwidth}{@{\extracolsep{\fill}}|l||r|r|r|} \hline
			Clutch\_Pedal = PRESSED & $\True$ & $\bullet$ & $\bullet$ \\ \hline
			Steering\_Wheel = USED & $\bullet$ & $\True$ & $\bullet$ \\ \hline
			Gearbox $\neq$ NEUTRAL & $\bullet$ & $\bullet$ & $\True$ \\ \hline
	\end{tabular*}
	\actionM{HMI\_Stop\_Ena}{FALSE}
	\conditionNS{$\neg d$}
	\actionMNS{HMI\_Stop\_Ena}{TRUE}
\end{spec}
\caption{RSML - Assignment specification for HMI\_Stop\_Ena}
\label{fig:rsml_assignment_spec}
\end{figure} 

The second concept is to define a state machine whose transitions are guarded with conditions on the input variables. The state machine serves as an abstraction on complicated conditions on input variables and is described using graphical state diagrams showing the states and transitions but not the conditions on the transitions. AND/OR tables are used again to specify the transition conditions.

\textbf{Summary} Applying RSML for the specification of automotive applications showed very promising results. We were able to express the complete functional behaviour of the Start/Stop System in RSML. The language was formal enough to describe precisely the functional behaviour yet still readable by engineers which was very important in order to permit domain engineers to validate the specification against the requirements document without needing special training in formal methods. However, we did not have tool support for RSML which was both an advantage and a disadvantage. The advantage of not having a tool was that it allowed us to have more freedom in the structure of the specification. The disadvantage was that we did not have the possibility of automatically checking the specification for consistency.

%
%
%
%
%
%
%

\subsection{Formal Modelling}
\label{subsec:method_formal_modelling}
\textbf{Constraints} There are two purposes of the formal modelling phase: One is to translate the specification into a formal model, i.e., a mathematically precise description of the functional behaviour of the system. The second purpose is to formalize the requirements in order to make them amenable for formal verification.\\ 
Such a formal model should provide the basis for formal verification. Thus, the formal language used for formal modelling must be formal enough to describe precisely the functional behaviour specified in the specification and to formalize the requirements we would like to prove on the formal model. In order to make these informal descriptions accessible to formal verification they have to be stated formally as well. Furthermore, the formal language must suit the application area
(i.e., description of state-based systems) 
and provide means for structuring the formal model. In addition to that, the formal language must be concrete enough to generate code from the formal model.\\
Event-B \cite{Abrial10} fulfills most of the constraints mentioned above. It is suited for the description of state-based systems since it is based on action transition systems and it is formal enough to describe precisely the functional behaviour as well as a large number of the properties we would like to prove about the system as invariants. Furthermore, it provides a refinement mechanism which allows us to start with an abstract formal model which can later be refined to a concrete model which provides the basis for code generation. 


\textbf{Description of method} Formal modelling in the language Event-B typically starts with a very abstract model which is refined step-by-step until the system and the environment has been completely modelled. For the Start/Stop System the formal modelling starts with a very abstract model containing only the output of the Start/Stop System. This model is then refined step-by-step. In each refinement step additional components described in the specification document are added to the formal model. Typed input and output variables of components described in the specification are modelled as \textit{variables} in the Event-B model. The types of these variables are specified using \textit{type invariants}. Each assignment specification and each transition of a state machine described in the specification is modelled by \textit{events} in Event-B, i.e., the conditions for the assignment are described as \textit{guards} of the event whereas the assignment itself is described using an \textit{action} of the event. It is important to note that the Event-B model also contains events for the system environment which models changes of system inputs. For example, the Event-B model for the Start/Stop System contains unguarded events modelling changes of input variables such as Clutch\_Pedal, Gearbox, and Steering\_Wheel.

\textbf{Example:} Figure \ref{fig:eventb_startstop} shows how the assignment specification for the output variable HMI\_Stop\_Ena in RSML (shown in Example 4) is translated into Event-B syntax. 


\begin{figure}[!htbp]\centering
$
{\small{
\begin{array}{@{}l@{}}
\mathsf{variables}\\ 
\quad \mathrm{HMI\_Stop\_Ena}\ \mathrm{Clutch\_Pedal}\\ 
\quad \mathrm{Gearbox}\ \mathrm{Steering\_Wheel}\\
\mathsf{invariants}\\
\quad \mathit{@inv1}\ \mathrm{HMI\_Stop\_Ena} \in \mathrm{BOOL}\\
\quad \mathit{@inv2}\ \mathrm{Clutch\_Pedal} \in \mathrm{T\_Clutch\_Pedal}\\
\quad \mathit{@inv3}\ \mathrm{Steering\_Wheel} \in \mathrm{T\_Steering\_Wheel}\\
\quad \mathit{@inv4}\ \mathrm{Gearbox} \in \mathrm{T\_Gearbox}\\
\mathsf{events}\\
\quad \mathsf{event}\ \mathrm{Set\_HMI\_Stop\_Ena\_FALSE}\\
\quad\quad \mathsf{when}\\ 
\quad\quad\quad \mathit{@grd1}\ \mathrm{Clutch\_Pedal = PRESSED\ \vee}\\ 
\quad\quad\quad \quad \quad	\quad	  \mathrm{Steering\_Wheel = USED\ \vee}\\
\quad\quad\quad \quad	\quad	\quad	  \mathrm{Gearbox \neq NEUTRAL}\\
\quad\quad \mathsf{then}\\
\quad\quad\quad \mathit{@act1}\ \mathrm{HMI\_Stop\_Ena := FALSE}\\ 
\quad \mathsf{event}\ \mathrm{Set\_HMI\_Stop\_Ena\_TRUE}\\
\quad\quad \mathsf{when}\\ 
\quad\quad\quad \mathit{@grd1}\ \mathrm{Clutch\_Pedal \neq PRESSED}\\ 
\quad\quad\quad \mathit{@grd2}\ \mathrm{Steering\_Wheel \neq USED}\\
\quad\quad\quad \mathit{@grd3}\ \mathrm{Gearbox = NEUTRAL}\\
\quad\quad \mathsf{then}\\
\quad\quad\quad \mathit{@act1}\ \mathrm{HMI\_Stop\_Ena := TRUE}\\ 
\quad \mathsf{end}\\
\mathsf{end}
\end{array}
}}
$
\caption{Event-B model for HMI\_Stop\_Ena}
\label{fig:eventb_startstop}
\end{figure}


As you can see in Figure \ref{fig:eventb_startstop} the output and input variables are modelled as Event-B variables. Their types
are specified by Event-B invariants. The assignment specification for the output variable $\mathrm{HMI\_Stop\_Ena}$ is modelled as two Event-B events depending
whether $\mathrm{HMI\_Stop\_Ena}$ is set to TRUE or FALSE.

The main feature of Event-B with which to state properties for a model is the concept of invariants. These invariants describe predicates that are proven to always hold.
Certain safety properties can be easily described as invariants
(e.g., if a defined output of the system is generally forbidden). An example from the Start/Stop System is that there should never be the request to start and the request to stop the engine at the same time. This kind of property is naturally suitable for formalization as invariants.\\

\textbf{Summary}
With Event-B and Rodin we were able to model the discrete part of our systems. Rodin has the great advantage of integrating the formal modelling phase and the formal verification phase so they can be treated in parallel --- this is important in helping to eliminate errors as soon as possible. 
Processes like configuration management, variant management, team development, version management etc. have to be better supported.
Scalability for industrial applications and more flexibility for decomposition and architecture have to be addressed in the future. For the formalization of requirements the concept of invariants in Event-B shows limitations. 

We had over 4000 generated proof obligations in the Start/Stop System, around 90\% of proof obligations were proven automatically by the provers integrated in Rodin. A large majority of the remaining manual proofs were very simple and might be proven automatically in the future with better adjustment and further development of the provers.

\section{Related Work}
\label{sec:related_work}
Costs and benefits of model-based development of embedded systems in the automotive industry have been examined in \cite{IGI2011}. The book chapter describes the results of a global study by Altran Technologies, the chair of software and systems engineering and the chair of Information Management of the Technical University of Munich. This work intends to cover a gap in research analyzing the status quo of model-based development and its effects on the economics. One of the authors of this work, Manfred Broy, has a vast literature on software engineering methods applied to the automotive sector, for example \cite{Broy06}. In \cite{Broy10} he presents a perspective which is very close to the one supported by our work. In his paper, Broy, discusses the need for a portfolio of models and methods and he emphasizes the importance of tool support. 

\section{Conclusions}
\label{sec:conclusions}

Formal methods are considered attractive by many researchers because concepts such as theorems, proof obligations, equations and others can be applied. However, academic attractiveness by itself does not justify industrial deployment. The work presented in this article shows  how elaborating a methodology based on a portfolio of different formalism, each tuned to a specific phase of development, allowed for a better set of requirements and, eventually, better code. Another criticism to FM is often based on the idea that specifications fulfilling the requirement of being interpreted formally are hard to write when compared with learning a new programming language. DEPLOY, and in particular the work presented here, actually demonstrated the opposite. On the other hand, the criticism that it is not possible to prove that formal methods can offer the same quality for less is still open, i.e., we have not empirically (numerically) shown that formal methods are cheaper. There is high confidence that the quality is better, but the added value is limited when the quality is already very good. 

This article discussed several software engineering issues, some of which are still open at present. The lack of a rigorous and repeatable approach of many "formal methods" significantly restricts the choice when it comes to identify a suitable formalism for a specific problem. In \cite{MazzaraDEPLOY09} this issue is historically investigated and the requirements of a "formal method" are identified to discover that many so-called ``methods'' are actually no more than notations, i.e., just formalisms without an attached rigorously defined and repeatable, systematic approach. Event-B is not one of those. Its refinement strategy has been demonstrated to be useful when applied to several case studies in a number of projects like RODIN \cite{RODINP} and DEPLOY itself \cite{DEPLOY}. However, not even Event-B is a panacea applicable to every phase of software development. In this article, we presented a strategy based on a formalism-based toolkit, i.e., a portfolio of formalisms where every specific phase of development has been attacked  by a different and suitable notation. The overall strategy proved to be a successful one and, given the thorough documentation generated by the project (\cite{DEPLOYD15}, \cite{DEPLOYD19}, \cite{DEPLOYD38}), it promises to be repeatable by engineers with an initially limited knowledge of formal methods. The importance of training here cannot be underestimated. 


The idea of this paper can be generalized in a way which sees software development as a sequence of stages with associated notations and techniques. Each phase should have an artifact or document as an input and will generate an adequate output. Given this broader framework, what presented in this paper should be considered just a specific instance for an application scenario (automotive). It is a matter of investigation (for which we do not have a full understanding at the moment) how the single (or multiple) notation(s) for each step should be chosen. We also need to understand what criteria inputs and outputs should individually follow and how they should be related to each other (for example in terms of pre and post conditions?). Even if our investigation necessarily leaves all this unsolved, we still believe it has clarified several aspects of industrial deployment of formal methods in automotive applications. 

\section*{Acknowledgment}
This work has been funded by the EU FP7 DEPLOY Project (Industrial deployment of system engineering methods providing high dependability and productivity). We would like to thank Cliff Jones, Alexander Romanovsky and John Fitzgerald for their valuable support.

\bibliographystyle{IEEEtran}
\bibliography{current-refs}

\begin{thebibliography}{10}
\providecommand{\url}[1]{#1}
\csname url@samestyle\endcsname
\providecommand{\newblock}{\relax}
\providecommand{\bibinfo}[2]{#2}
\providecommand{\BIBentrySTDinterwordspacing}{\spaceskip=0pt\relax}
\providecommand{\BIBentryALTinterwordstretchfactor}{4}
\providecommand{\BIBentryALTinterwordspacing}{\spaceskip=\fontdimen2\font plus
\BIBentryALTinterwordstretchfactor\fontdimen3\font minus
  \fontdimen4\font\relax}
\providecommand{\BIBforeignlanguage}[2]{{%
\expandafter\ifx\csname l@#1\endcsname\relax
\typeout{** WARNING: IEEEtran.bst: No hyphenation pattern has been}%
\typeout{** loaded for the language `#1'. Using the pattern for}%
\typeout{** the default language instead.}%
\else
\language=\csname l@#1\endcsname
\fi
#2}}
\providecommand{\BIBdecl}{\relax}
\BIBdecl

\bibitem{DEPLOY}
\BIBentryALTinterwordspacing
``{DEPLOY}: {I}ndustrial deployment of system engineering methods providing
  high dependability and productivity.'' [Online]. Available:
  \url{http://www.deploy-project.eu/}
\BIBentrySTDinterwordspacing

\bibitem{EventBBook}
J.-R. Abrial, \emph{Modelling in Event-B}.\hskip 1em plus 0.5em minus
  0.4em\relax Cambridge University Press, 2010.

\bibitem{RodinPlatform}
{The RODIN platform}, online at http://rodin-b-sharp.sourceforge.net/.

\bibitem{RODIN}
\BIBentryALTinterwordspacing
``{E}vent-{B} and the {R}odin {P}latform.'' [Online]. Available:
  \url{http://www.event-b.org/}
\BIBentrySTDinterwordspacing

\bibitem{AbrialZ-book}
J.-R. Abrial, S.~A. Schuman, and B.~Meyer, \emph{A Specification
  Language}.\hskip 1em plus 0.5em minus 0.4em\relax New York, NY, USA:
  Cambridge University Press, 1980.

\bibitem{AbrialB-book}
J.-R. Abrial, \emph{{T}he {B}-{B}ook: Assigning programs to meanings}.\hskip
  1em plus 0.5em minus 0.4em\relax New York, NY, USA: Cambridge University
  Press, 1996.

\bibitem{Abrial10}
------, \emph{{T}he {E}vent-{B} {B}ook}.\hskip 1em plus 0.5em minus 0.4em\relax
  Cambridge, UK: Cambridge University Press, 2010.

\bibitem{BjornerVDM78}
\BIBentryALTinterwordspacing
D.~Bj{\o}rner and C.~B. Jones, Eds., \emph{The Vienna Development Method: The
  Meta-Language}, ser. Lecture Notes in Computer Science.\hskip 1em plus 0.5em
  minus 0.4em\relax Springer-Verlag, 1978, vol.~61. [Online]. Available:
  \url{https://www.springerlink.com/content/ql766633l472/}
\BIBentrySTDinterwordspacing

\bibitem{Jones80a}
\BIBentryALTinterwordspacing
C.~B. Jones, \emph{Software Development: A Rigorous Approach}.\hskip 1em plus
  0.5em minus 0.4em\relax Englewood Cliffs, N.J., USA: Prentice Hall
  International, 1980. [Online]. Available:
  \url{http://portal.acm.org/citation.cfm?id=539771}
\BIBentrySTDinterwordspacing

\bibitem{Jones90a}
\BIBentryALTinterwordspacing
------, \emph{Systematic Software Development using VDM}, 2nd~ed.\hskip 1em
  plus 0.5em minus 0.4em\relax Prentice Hall International, 1990. [Online].
  Available:
  \url{http://homepages.cs.ncl.ac.uk/cliff.jones/ftp-stuff/Jones1990.pdf}
\BIBentrySTDinterwordspacing

\bibitem{Mazzara10}
M.~Mazzara and A.~Bhattacharyya, ``On modelling and analysis of dynamic
  reconfiguration of dependable real-time systems,'' in \emph{DEPEND,
  International Conference on Dependability}, 2010.

\bibitem{MazzaraDEPLOY09}
M.~Mazzara, ``Deriving specifications of dependable systems: toward a method,''
  in \emph{Proceedings of the 12th European Workshop on Dependable Computing
  (EWDC)}, 2009.

\bibitem{Mazzara11}
------, ``On methods for the formal specification of fault tolerant systems,''
  in \emph{DEPEND, International Conference on Dependability}, 2011.

\bibitem{WoodcockEt09}
J.~Woodcock, P.~G. Larsen, J.~Bicarregui, and J.~Fitzgerald, ``{F}ormal
  {M}ethods: {P}ractice and {E}xperience,'' \emph{ACM Computing Surveys},
  vol.~41, no.~4, Oct 2009.

\bibitem{Jackson00}
M.~Jackson, \emph{Problem Frames: Analyzing and structuring software
  development problems}.\hskip 1em plus 0.5em minus 0.4em\relax Addison-Wesley,
  2000.

\bibitem{Leveson:1994}
\BIBentryALTinterwordspacing
N.~G. Leveson, M.~P.~E. Heimdahl, H.~Hildreth, and J.~D. Reese, ``Requirements
  {S}pecification for {P}rocess-{C}ontrol {S}ystems,'' \emph{IEEE Trans. Softw.
  Eng.}, vol.~20, pp. 684--707, September 1994. [Online]. Available:
  \url{http://portal.acm.org/citation.cfm?id=188229.188234}
\BIBentrySTDinterwordspacing

\bibitem{Heimdahl:1996}
M.~Heimdahl and N.~Leveson, ``Completeness and consistency in hierarchical
  state-based requirements,'' \emph{Software Engineering, IEEE Transactions
  on}, vol.~22, no.~6, pp. 363 --377, jun 1996.

\bibitem{Harel:1987}
\BIBentryALTinterwordspacing
D.~Harel, ``Statecharts: A visual formalism for complex systems,'' \emph{Sci.
  Comput. Program.}, vol.~8, pp. 231--274, June 1987. [Online]. Available:
  \url{http://dl.acm.org/citation.cfm?id=34884.34886}
\BIBentrySTDinterwordspacing

\bibitem{BBook}
J.-R. Abrial, \emph{The B-Book}.\hskip 1em plus 0.5em minus 0.4em\relax
  Cambridge University Press, 1996.

\bibitem{DEPLOYD38}
K.~Grau, R.~Gmehlich, F.~Loesch, J.-C. Deprez, R.~D. Landtsheer, and
  C.~Ponsard, ``{DEPLOY} {D}eliverable {D38}: {R}eport on {E}nhanced
  {D}eployment in the {A}utomotive {S}ector,'' DEPLOY Project, Tech. Rep. D38,
  2011.

\bibitem{IGI2011}
M.~Broy, S.~Kirstan, H.~Krcmar, B.~Schaetz, and J.~Zimmermann, ``What is the
  benefit of a model-based design of embedded software systems in the car
  industry?'' in \emph{Emerging Technologies for the Evolution and Maintenance
  of Software Models}.\hskip 1em plus 0.5em minus 0.4em\relax IGI Global, 2011,
  pp. 410--443.

\bibitem{Broy06}
M.~Broy, ``Challenges in automotive software engineering.'' in \emph{ICSE'06},
  2006, pp. 33--42.

\bibitem{Broy10}
------, ``Seamless method- and model-based software and systems engineering,''
  in \emph{The Future of Software Engineering}.\hskip 1em plus 0.5em minus
  0.4em\relax Springer, 2010, pp. 33--47.

\bibitem{RODINP}
\BIBentryALTinterwordspacing
``{RODIN}: {R}igorous {O}pen {D}evelopment {E}nvironment for {C}omplex
  {S}ystems.'' [Online]. Available: \url{http://rodin.cs.ncl.ac.uk/}
\BIBentrySTDinterwordspacing

\bibitem{DEPLOYD15}
C.~Jones, ``{DEPLOY} {D}eliverable {D15}: {A}dvances in {M}ethodological
  {WP}s,'' http://www.deploy-project.eu/pdf/D15final.pdf, Tech. Rep. D15, 2009.

\bibitem{DEPLOYD19}
F.~Loesch, R.~Gmehlich, K.~Grau, M.~Mazzara, and C.~Jones, ``{DEPLOY}
  {D}eliverable {D19}: {P}ilot {D}eployment in the {A}utomotive {S}ector,''
  DEPLOY Project, Tech. Rep. D19, 2010.

\end{thebibliography}

\end{document}